\documentclass[11pt]{article}
\usepackage{amsmath,amsthm,amssymb}
\usepackage{graphics}
\usepackage{psfrag}
\usepackage[letterpaper,margin=1.2in]{geometry}
\usepackage[utf8]{inputenc}
\usepackage[pdftex]{graphicx}	% using pdflatex

\def\R{\mathbb{R}}

\def\E{\mathbb{E}}

\newcommand{\be}{\begin{equation}}
\newcommand{\ee}{\end{equation}}
\newcommand{\bea}{\begin{eqnarray}}
\newcommand{\eea}{\end{eqnarray}}
\newcommand{\beann}{\begin{eqnarray*}}
\newcommand{\eeann}{\end{eqnarray*}}
\newcommand{\benn}{\begin{equation*}}
\newcommand{\eenn}{\end{equation*}}

\usepackage{todonotes}

\usepackage{soul}
\usepackage{color}

\newcommand{\cO}{{\mathcal O}}  % calligraphic O
\def\ra{\rightarrow}

\begin{document}
% Start the document...

\author{Christian Kuehn\footnotemark[1]~\footnotemark[5]~ and Erik A. Martens\footnotemark[2]~\footnotemark[3]~\footnotemark[5]~ and Daniel M. Romero\footnotemark[4]}

\renewcommand{\thefootnote}{\fnsymbol{footnote}}
\footnotetext[1]{%
Institute for Analysis and Scientific Computing, 
Vienna University of Technology, 
1040 Vienna, Austria.} 
\renewcommand{\thefootnote}{\arabic{footnote}}

\renewcommand{\thefootnote}{\fnsymbol{footnote}}
\footnotetext[2]{%
Max Planck Institute for Dynamics and Self-Organization, 
37077 Göttingen, Germany.} 
\renewcommand{\thefootnote}{\arabic{footnote}}

\renewcommand{\thefootnote}{\fnsymbol{footnote}}
\footnotetext[3]{%
Department of Biomedical Sciences, 
Copenhagen University, 
Blegdamsvej 3, 2200 Copenhagen, Denmark}
\renewcommand{\thefootnote}{\arabic{footnote}}

\renewcommand{\thefootnote}{\fnsymbol{footnote}}
\footnotetext[4]{%
Northwestern University Institute on Complex Systems,
Evanston IL 60208, USA.} 
\renewcommand{\thefootnote}{\arabic{footnote}}

\renewcommand{\thefootnote}{\fnsymbol{footnote}}
\footnotetext[5]{%
equal contribution} 
\renewcommand{\thefootnote}{\arabic{footnote}}

\title{Critical Transitions in Social Network Activity}

\maketitle

\begin{abstract}
A large variety of complex systems in ecology, climate science, biomedicine
and engineering have been observed to exhibit tipping points, where the internal
dynamical state of the system abruptly changes. For example, such critical
transitions may result in the sudden change of ecological environments and
climate conditions. Data and models suggest that detectable warning signs
may precede some of these drastic events. This view is also corroborated
by abstract mathematical theory for generic bifurcations in stochastic
multi-scale systems. Whether the stochastic scaling laws used as warning
signs are also present in social networks that anticipate a-priori {\it unknown}
events in society is an exciting open problem, to which at present only highly
speculative answers can be given. Here, we instead provide a first step towards
tackling this formidable question by focusing on a-priori {\it known} events
and analyzing a social network data set with a focus on classical variance and
autocorrelation warning signs. Our results thus pertain to one absolutely
fundamental question: Can the stochastic warning signs known from other areas
also be detected in large-scale social network data? We answer this question
affirmatively as we find that several a-priori known events are preceded by variance
and autocorrelation growth. Our findings thus clearly establish the necessary
starting point to further investigate the relation between abstract mathematical
theory and various classes of critical transitions in social networks. 

\end{abstract}
\section{Introduction}

Can sudden changes in society be anticipated by monitoring social network activity?
This problem has recently gained considerable attention and serves as the background
motivation for our work. It has been suggested \cite{Cohen,Webster} in the media
that certain societal-scale oppositional movement may be preceded and even be
triggered by social network activity and that those events could be viewed as
'tipping points' \cite{HudsonFlannes}. For example, the events in Egypt in 2011
have been in the focus of recent research~\cite{Lim,TufekciWilson,Lotanetal} which
has a similar conjecture (\textit{``Tahrir Square was a foreseeable surprise:
tracing the history of Egyptian activism''}~\cite{Lim1}). To address such a
conjecture from a theoretical modeling perspective would require a deep
understanding of how new social media impact collective action~\cite{BimberFlanaginStohl}
and how cyber-collective movements are formed~\cite{AgarwalLimWigand}. A
conclusive study to confirm the conjecture that the a-priori unknown revolution
in Egypt in 2011 was foreseeable is far beyond the reach of current research. 
Hence, while we cannot directly solve this difficult problem directly, we must
reduce it to several elementary (and thus more tractable) problems concerning
complex network theory. One such necessary first step towards solving this larger
problem is to ask: may the well-known warning signs preceding drastic events
in social media be found at all? In this study, we answer this question by analyzing
social media data for warning signs in the context of well-defined a-priori known
events such as special public holidays. 

In particular, our approach establishes a first potential link between social media
analysis and the recent theory of warning signs for critical transitions (or tipping
points) \cite{Schefferetal,KuehnCT1}. Similar approaches have received major recent
attention in ecology where warning signs were detected in experimental
\cite{DaiVorselenKorolevGore,DrakeGriffen,Veraartetal} and in field data
\cite{Wangetal,Carpenteretal1}. A critical transition may informally be defined as a
rapid drastic change of a time-dependent dynamical system; for more precise definitions
see \cite{KuehnCT2,AshwinWieczorekVitoloCox}. Warning signs for critical transitions
have been investigated intensively in ecological models during the last decade
\cite{Dakosetal1,LadeGross}. Similar results have also been obtained in the context of
climate science~\cite{Lenton,ThompsonSieber2,CimatoribusDrijfhoutLivinavanderSchrier},
biomedical applications \cite{Venegasetal,MeiselKuehn}, engineering \cite{Chertkovetal}
and epidemiology \cite{KuehnCT2,OReganDrake}. These studies led to the conjecture that
there are some warning signs which are generic \cite{Schefferetal} for large classes of
natural systems. From a mathematical perspective, this conjecture can be made precise
for transitions near certain bifurcation points; see \cite{KuehnCT1,KuehnCT2} and
Appendix \ref{sec:methods1}. Two of the most classical warning signs are rising variance
and rising auto-correlation before a critical transition \cite{CarpenterBrock,Schefferetal}
whose origin is described in more detail in Appendix \ref{sec:methods1}. The basic idea
is that if a drastic change is induced by a critical (bifurcation) point, then the
underlying deterministic dynamics becomes less stable. Hence, the noisy fluctuations
become more dominant as the decay rate decreases close to the critical transition.
As a result, (a) the variance in the signal increases, due to the stronger fluctuations
and (b) the system's state memory ({i.e.}, auto-correlation) increases due to smaller
deterministic contraction onto a single state \cite{Schefferetal,KuehnCT1}. It can be
shown that both warning signs are related via a suitable fluctuation-dissipation relation
\cite{DitlevsenJohnsen}. 

For social networks, the situation is much less developed. Although the notion of
'tipping' is somewhat familiar in sociological contexts \cite{Gladwell, Lamberson},
work on detailed statistical analysis of warnings in social networks from a dynamical
systems perspective is very sparse. One approach which is related and complementary
to the results presented in this paper is recent work by Slater~\cite{Slater} where a
data set of approximately 11000 blog posts is analyzed and tipping is defined using a
sentiment score. Based on this score, two special points were identified and warnings were
computed. In this paper, we analyze a large-scale social network data set and focus on
clearly defined events, which are well-localized in time.

\section{Results}
\label{sec:results}

The data set we analyzed consists of messages communicated publicly via Twitter. In
this context, a message is also called a tweet. The messages collected account for
about 20-30 percent of all the data tweeted world-wide in the time period from June
1 2009 to Dec 31 2009, amounting to 476,547,774 tweets, or on average over the time
period 92,767 tweets/hour \cite{YangLeskovec}; more details about the data set can be
found in Appendix \ref{sec:methods}.

We extract time series by counting word frequencies. The frequency count is based on
hashtags such as~\verb|#halloween|. Twitter users frequently attach hashtags to particular
events or topics \cite{RomeroMeederKleinberg}. For the words we consider, we remove all
white spaces and transform strings to lower case, {e.g.},~\verb|#halloween| equals
\verb|#Halloween|. Subsequently, linear trends were removed from the resulting time
series; for more details on the computation see Appendix \ref{sec:methods}.

\begin{figure}
\includegraphics[width=\textwidth]{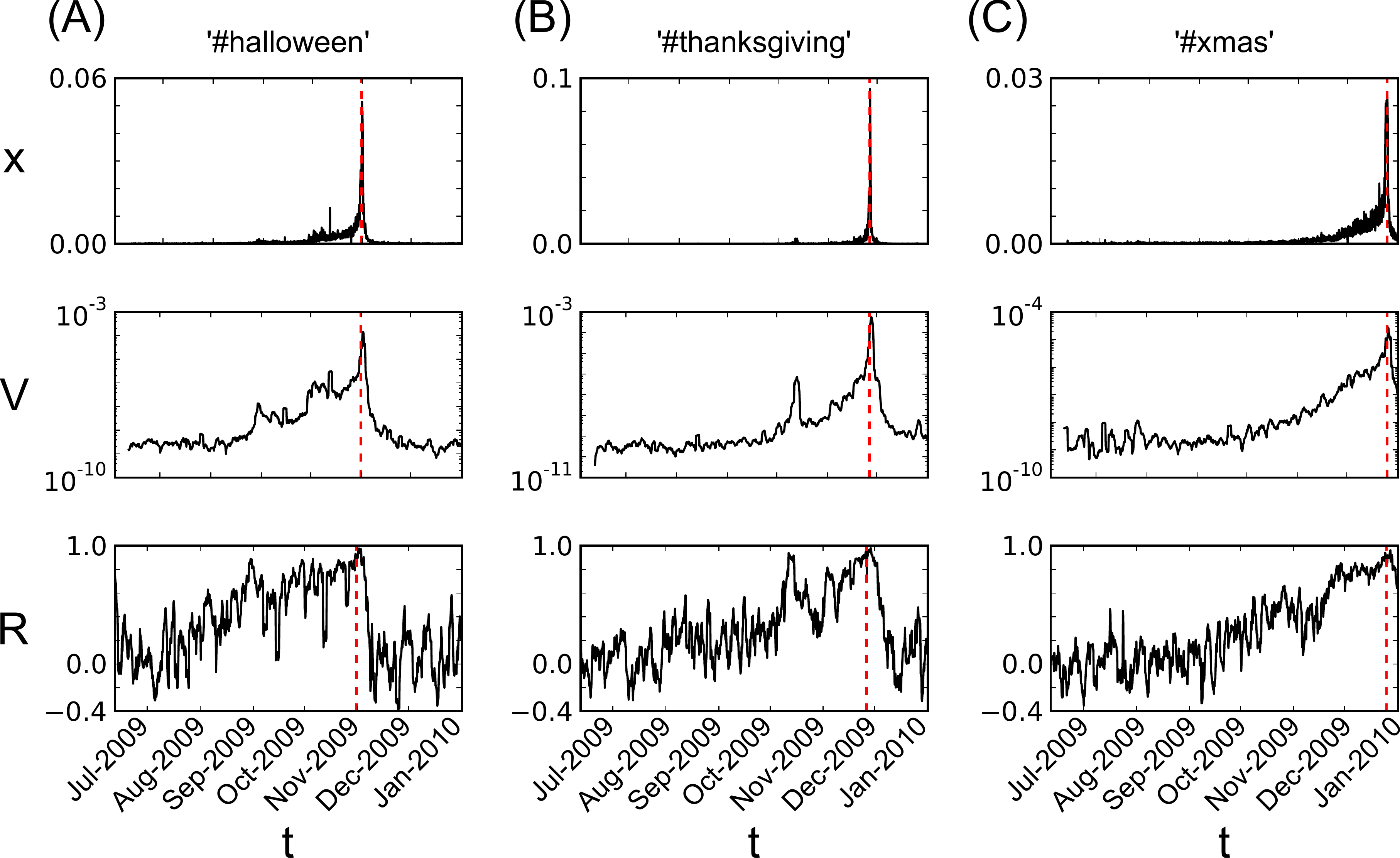}
\caption{\label{fig1} Hashtag frequency time-series reveal signatures of critical
transitions. (A) \#halloween (max. 6019 tweets/h) (B) \#thanksgiving ({max.}~10157
tweets/h) (C) \#xmas ({max.}~3501 tweets/h). Columns from top to bottom display
normalized word frequency $x$, variance $V$ and lag-1 autocorrelation $R$ calculated
from the time series of word frequency. The red vertical dashed line indicates the
a-priori known event date.}
\end{figure}

We have already pointed out that it is difficult to analyze \textit{a-priori unknown
events}, such as the revolution in Egypt in 2011, using social network data. Such
events are not well localized in time, so it is hard to even determine when a critical
transition happens. However, one may pose the simpler question of whether the scaling
laws for the variance and autocorrelation warning signs can be detected in the time
period preceding an \textit{a-priori known event}. This is a test of whether it may be
possible to extract scaling laws from large-scale social networks near critical
transitions at all. Since our data set encompasses fall/winter 2009 we chose the three
special events: 

\begin{itemize}
 \item[(A)] \verb|#halloween|: Halloween occurred on Saturday, October 31st, 2009
 \item[(B)] \verb|#thanksgiving|: Thanksgiving Day occurred on Thursday, November 26th, 2009
 \item[(C)] \verb|#xmas|: Christmas Day occurred on Friday, December 25th, 2009
\end{itemize}

Note that these events are well-defined in the sense that the event dates are fixed
and the hashtags are directly associated to the events since they do not commonly occur
in different contexts. The first row of Figure \ref{fig1} shows the time series of frequency
for each hashtag. As expected, the overall hashtag frequency count starts to increase weeks
before the events (A)-(C). Note that we do know \textit{a-priori} that a drastic change is
going to occur at the precise event date. However, if we would terminate the time series
several days or weeks before the event and did not know about (A)-(C) then we would have
to take into account the possibility that a drastic spike never occurs and the word frequency
just decreases slowly.

From this last perspective, {i.e.}, by using only past data and information, we
calculated the variance and lag-1 autocorrelation by considering sliding averages over
a period of 50 hours and time steps of 5 hours. The results are shown in second and
third rows of Figure~\ref{fig1} respectively. In all three cases, we see a clear increase
of the variance. The autocorrelation also increases before the events (A)-(C), as shown
in row three of Figure~\ref{fig1}. Again, the warning sign is clearly visible for all three
cases.

\begin{figure}
\includegraphics[width=\textwidth]{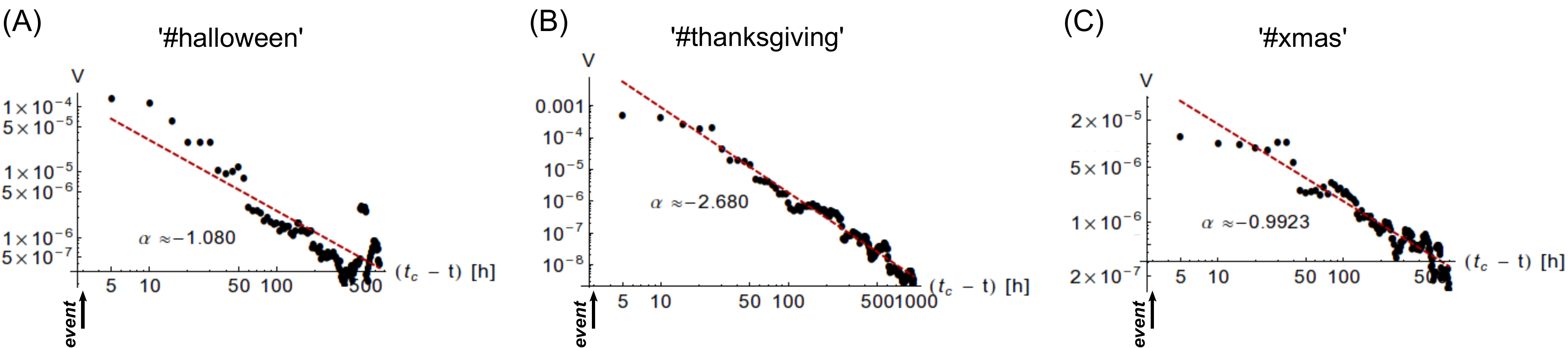}
\caption{\label{fig3}
Scaling exponents $\alpha$ from \eqref{eq:scaling_law} extracted from the time
series data using least squares fit (dashed red line) for the three hashtags
from Figure~\ref{fig1}. Data is shown on a semilog scale with reverse time on
the horizontal axis, and events are marked with arrows.}
\end{figure}

Furthermore, we investigated the scaling laws for the variance. The mathematical
theory of critical transitions indicates that there is a regime in which we may
find a scaling law of the form
\be
\label{eq:scaling_law}
\text{Var($t$)}=A(t_c-t)^\alpha, \qquad \text{as $t\ra t_c$},
\ee
where $t_c$ denotes the time of the spiking event and where $A>0$, $\alpha<0$
are two constants. More details on the theory behind the power law scaling
\eqref{eq:scaling_law} is provided in Appendix \ref{sec:methods1}; however, we
mention here that there are generically three different regimes for stochastic
scaling laws: (R1) far away from the bifurcation point, (R2) a large-in-time
regime approaching the bifurcation point and (R3) a small-in-time regime near
the bifurcation. Theory predicts that a scaling law \eqref{eq:scaling_law} should
hold in regime (R2) and that the key measure is the exponent $\alpha<0$ as it
can potentially distinguish between different dynamical transitions. Figure
\ref{fig3} shows the results for exponent $\alpha$ for the three cases from
Figure \ref{fig1}. We find that there is indeed a regime distinction between
(R1), (R2) and (R3). We did perform a least-squares fit to determine $\alpha$
in (R2). Cases (A) and (C) show an exponent very close to $\alpha=-1$ which 
is an exponent generically expected near several types of bifurcation points
such as Hopf, transcritical and pitchfork bifurcations. However, (B) shows a
substantially smaller exponent than predicted by theoretical considerations. There
could be several potential explanations for this non-standard exponent;
see Appendix \ref{sec:methods1}. Nevertheless, in all three cases we see a
clear power law scaling behavior of the variance that is characteristic of the
regime type (R2).

As a last aspect of our work, we also analyzed the Twitter time series data
to find examples of other events that are, {e.g.}, unpredictable,
exogenously-induced, periodically-driven, with short periods of warning
signs or highly noise-driven. Several alternative examples of such complementary
behaviors are presented in Appendix~\ref{sec:further}. These time series illustrate
that there is a variety of potentially novel dynamical behaviors in large-scale
social networks near large spikes that deserve to be investigated in their own right.   

\section{Discussion}
\label{sec:discussion}

We have demonstrated that clear warning signs exist for a-priori known events
in large-scale social media data. Our results leave open the possibility that
warning signs could also exist for \textit{a-priori unknown} events. Investigating
this problem further still remains a largely unexplored research topic. In particular,
it is important to check various conjectures on the influence of new social media
on major societal events very carefully.

Even very elementary questions are open in the context of social network data
and warning signs. For instance, what type of time series data should be used?
In this study, we analyzed the content of messages posted by users, but one could
equally well think of analyzing the dynamics of connections made between users,
or various other network activities. Even when considering the content of messages,
options other than the topic or hashtag considered here are conceivable, such as
particular words, word combinations, or even entire phrases. Further important
questions include:

\begin{itemize}
 \item How do we define when a critical transition occurs for an a-priori unknown
 event in the data?
 \item Can we link warning signs in social networks to a-priori unknown critical
 transitions outside a social network?
 \item Which models of social networks can re-produce critical transitions observed
 in data?
\end{itemize}

In summary, our contribution is to provide a proof-of-principle. In particular,
we have found stochastic scaling laws, which are characteristic signs of critical
transitions, in a large social media data set preceding large increases of activity.
This suggests that the underlying social network may undergo a qualitative change
between two different dynamic regimes. We explicitly emphasize that we neither
claim to link these signs to any particular underlying mathematical network model
nor can we prove for which types of network events such warning signs may exist;
however, we have shown that there exist signatures that do resemble much the structures
of warning signs observed in other fields, such as ecology, as well as the theoretical
scaling laws, thus opening new perspectives on research on social network dynamics.  

It is important to develop additional mathematical tools to answer further research
questions. Due to their present lack, we have chosen examples of events that are
sufficiently isolated and where influences that we cannot quantify are eliminated.
Warning signs in social network data are still a largely unexplored topic. We expect
that our work will stimulate further research, with ample opportunities to uncover
results on social network dynamics with far-reaching ramifications.\medskip

\textbf{Acknowledgments:} CK would like to thank the Austrian Academy of Sciences
(\"{O}AW) for support via an APART fellowship and the European Commission (EC/REA)
for support by a Marie-Curie International Re-integration Grant. EM's contribution
is part of the Dynamical Systems Interdisciplinary Network, University of Copenhagen.
DR acknowledges the generous research support sponsored by the Army Research
Laboratory and accomplished under Cooperative Agreement Number W911NF-09-2-0053
and DARPA BAA-11-64, Social Media in Strategic Communication. The views and
conclusions contained in this document are those of the authors and should not
be interpreted as representing the official policies, either expressed or implied,
of the Army Research Laboratory or the {U.S.}~Government. The {U.S.}~Government
is authorized to reproduce and distribute reprints for Government purposes
notwithstanding any copyright notation here on. 

\appendix

\section{Time Series Analysis}
\label{sec:methods}

We briefly describe the time series analysis. The basic time unit we use is the
number of hours elapsed since the starting point of the time series. Denote the
word frequency during one hour $x_t$. For example, for \verb|#xmas| we have
\begin{equation*}
x_t=\frac{\text{number of occurrence of \texttt{\#xmas} during [t-1,t]}}{{\text{tweet volume during [t-1,t]}}{}}
\end{equation*}
which yields a time series $\{x_1,x_2,\ldots,x_t,\ldots\}$. To compute
the warning signs we fix a window size $n$. At a time point $t_k> n+l$
for some fixed lag $l\geq 1$ we consider the vector
\begin{equation*}
{\bf x_k}:=(x_{t_k-(n-1)},x_{t_k-(n-2)},\ldots,x_{t_k}).
\end{equation*}
Denote the vector obtained by removing the linear trend from ${\bf x_k}$
by ${\bf y_k}$. Then we may just compute the mean $\mu_k$, variance $V_k$
and lag-$l$ autocorrelation $R_{k,l}$ of ${\bf y_k}$ by the standard formulas 
\begin{eqnarray*}
\mu_k&=&\frac{1}{n}\sum_{j=1}^n (y_k)_j,\quad V_k=\frac{1}{n}\sum_{j=1}^n ((y_k)_j-\mu_k)^2,\\
R_{k,l}&=&\frac{1}{(n-l)V_k}\sum_{j=1}^{n-l}(y_j-\mu_k)(y_{j+l}-\mu_k)
\end{eqnarray*}
This yields the required time series for the warning signs given by $V_k$
and $R_{k,l}$. Note that we also choose a time step $s$ for the spacing in
the index $k$ which yields that 
\begin{equation*}
{\bf V}=(\ldots,V_{k},V_{k+s},V_{k+2s},\ldots)
\end{equation*}
and similarly for ${\bf R_{l}}$. To understand why the variance and
autocorrelation are expected to increase generically near certain types
of critical transitions we refer the reader to \cite{KuehnCT1,Schefferetal,KuehnCT2}
where also the specific scaling laws are discussed.

\begin{figure}
\centering
  \includegraphics[width=0.5\textwidth]{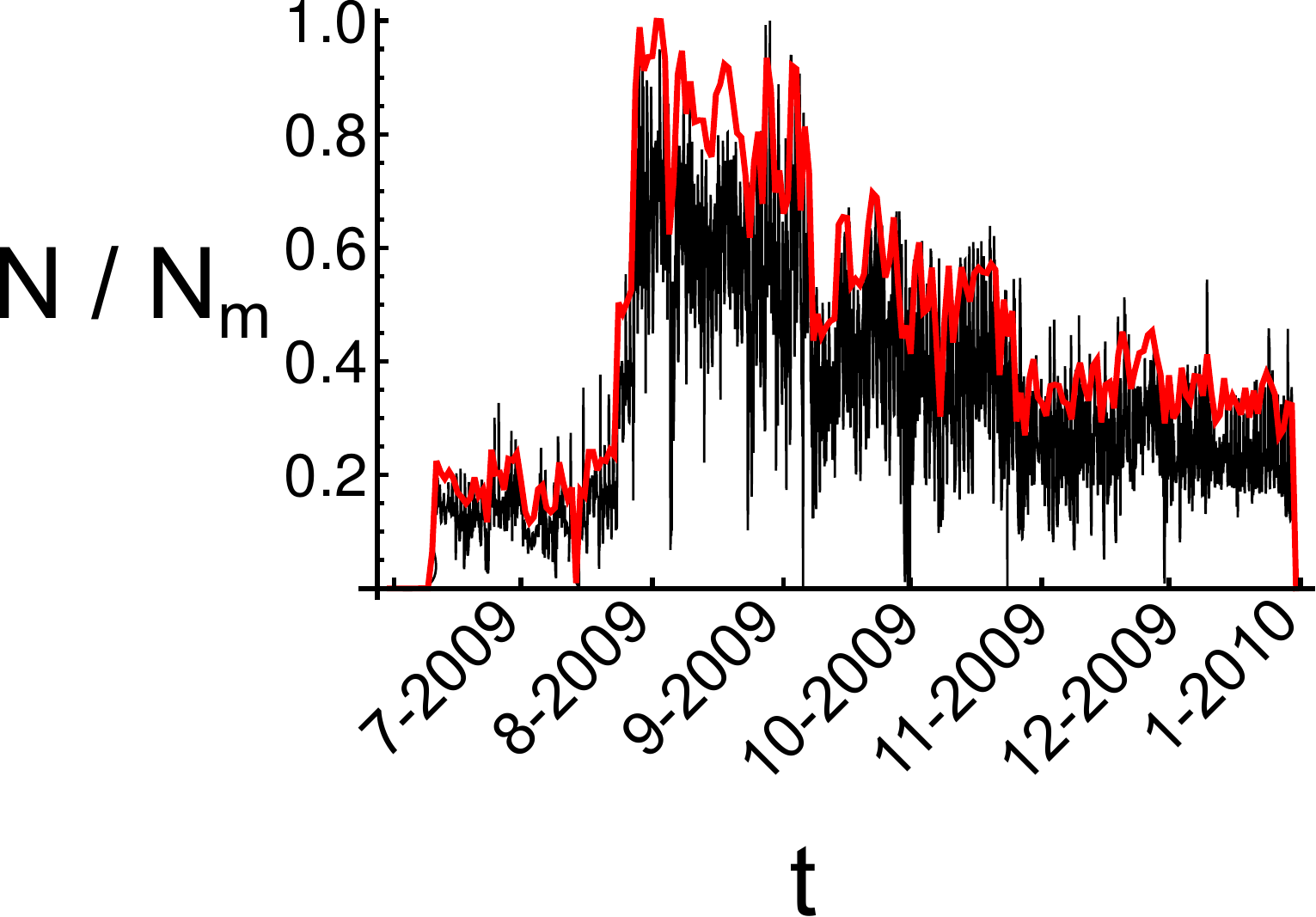}
  \caption{\label{figure2}Relative tweet volume $N/N_{max}$ per 
  hour (black) and per day (red) during the period from June 1 2009 
  to December 31 2009. }
\end{figure}

The relative tweet volume $N/N_m$ per hour and day are shown in Figure \ref{figure2}; 
the maximum recorded volume is $N_m=5,106,720$ / hour or $N_m=296,256$ / day, 
respectively. The average tweet activity is $2.2165\times10^6$ / day or $92767.7$ / hour,
with  a standard deviation of $1.27554\times10^6$ / day or $59375.9$ / hour, respectively.

\section{Critical Transitions / Tipping Points}
\label{sec:methods1}

In this section we provide some technical background on critical transitions
(or tipping points) to support our main theme of studying variance and autocorrelation
near jumps in time series of complex networks. Consider a general $n$-dimensional
system of ordinary differential equations (ODEs) given by
\be
\label{eq:ODE}
\frac{dx}{dt}=f(x,\tilde{\mu}),\qquad (x,\tilde{\mu})\in\R^m\times \R^p
\ee  
where we may view $\tilde{\mu}$ as the parameters describing the network
properties and $x$ as all the relevant variables to track the state of the
network; the map $f$ is assumed to be sufficiently smooth. Suppose there
exists a branch of equilibria  $x^*=x^*(\tilde{\mu})$, {i.e.}, steady states
with  $f(x^*(\tilde{\mu}),\tilde{\mu})=0$. Suppose the equilibria are stable
for a certain range of parameter values and lose stability for some
$\tilde{\mu}=\tilde{\mu}^*$ at which the state of the system ({i.e.}~the
network) changes drastically. In this case it can be proven rigorously that
the only two situations which imply stability loss and are stable to arbitrary
perturbations within the class of  sufficiently smooth vector fields upon
variation of a \emph{single} parameter are the so-called fold and Hopf
bifurcations (see the description of such (codimension one) bifurcations
in \cite[Ch.~3]{Kuznetsov} as well as \cite[Sec.~3.4]{GH}). Due to this theory
we may start from the low-dimensional cases $(m,p)=(1,1)$ for the fold
and $(m,p)=(2,1)$ for the Hopf bifurcation. The fold is always a tipping
point while the Hopf bifurcation only induces tipping to a distant attractor
in the subcritical case \cite[Prop.~2.5-2.6]{KuehnCT1}. We remark that one can
show that it is a generic property \cite[Sec.~7.4]{KuehnCT2} that the warning
signs we are going to discuss appear in all variables $x\in \R^m$ upon sufficient
coupling {i.e.}~before reducing to the minimal dimensions \cite[Sec.~3.2]{GH}.\medskip

For the fold bifurcation one may prove \cite[Sec.~3.2]{Kuznetsov} that the
system near the transition, where a stable and unstable steady state collide
and annihilate, may be reduced to
\be
\label{eq:fold}
\frac{dx}{dt}=\mu-x^2+\cO(\cdots),\qquad (x,\mu)\in\R\times \R
\ee   
where $\cO(\cdots)$ denotes higher-order terms which we shall drop from now
on and, for concreteness, we may think of $x$ as the word frequency. The
two steady states are $x^\pm=\pm\sqrt\mu$ for $\mu>0$ and $x^+$ is stable
while $x^-$ is unstable. The fold bifurcation occurs for $\mu=0$. Abstractly,
one may view the stable steady state $x^+$ as the representation of the word
frequency in the network before it approaches the drastic jump upon parameter
variation of $\mu$ (which we could just take as progression of time for our
case). Linearizing around $x^+$ yields the variational equation 
\be
\label{eq:variational}
\frac{dX}{dt}=(D_xf)(x^+)X=-2\sqrt\mu X\quad \Rightarrow X(t)=X(0)e^{-2\sqrt\mu t}.
\ee
From \eqref{eq:variational} it follows that the local linear exponential
stability of $x^+$ decreases as $\mu\ra 0$; this is just the classical
slowing-down phenomenon (or intermittency) \cite[p.343-346]{GH}. If one
perturbs \eqref{eq:fold} by an additive noise process $\xi(t)$ which is
delta-correlated white noise, {i.e.}, $\E[\xi(t)]=0$, 
$\E[\xi(t_1)\xi(t_2)]=\delta(t_1-t_2)$, and where $\sigma>0$ is a parameter
controlling the noise level, then this yields the stochastic differential
equation
\be
\label{eq:SDE}
\frac{dx}{dt}=\mu-x^2+\sigma ~\xi(t),
\ee
which is to be interpreted in the sense of the It\^{o}-calculus. Under
certain assumptions on the parameter variation of $\mu$ and the sufficiently
small noise level $\sigma$~\cite[p.~479]{KuehnCT2} one may prove that there
is a region before the fold where the time series fluctuating near the
deterministic steady-state exhibits a stochastic scaling law for the
variance
\be
\label{eq:res_fold}
\text{Var}(x(t))=\sigma^2 \cO\left(\frac{1}{\sqrt{\mu}}\right)+
\text{higher-order error terms}\sim\mathcal{O}\left(\frac{1}{\sqrt{\mu}}\right)
\ee 
as $\mu\ra 0$ from above and $\sigma>0$ remains fixed. A related result,
obtained via a fluctuation-dissipation relation, is that the autocorrelation
increases as $\mu\ra 0$ from above; see~\cite{DitlevsenJohnsen,Schefferetal}
for a discussion of this aspect. The argument for the case of the Hopf
bifurcation follows in an analogous manner and the resulting scaling law
turns out to be~\cite[Thm.~5.2]{KuehnCT2} 
\be
\label{eq:res_Hopf}
\text{Var}(x(t))=\sigma^2 \cO\left(\frac{1}{\mu}\right)+
\text{higher-order error terms}\sim\mathcal{O}\left(\frac{1}{\mu}\right).
\ee 
Hence, this motivates to test for an increase in variance and autocorrelation
in the time series data. Furthermore, replacing the parameter $\mu$ by
a time-dependent drift, or just by time $t$ for simplicity, we expect to
detect power law scalings of the form 
\be
\text{Var}(x(t))\sim A(t_c-t)^{\alpha},\qquad t\ra t_c,
\ee
where $t_c$ is the time when the critical transition occurs. The two
generic scaling exponents are $\alpha=-1$ and $\alpha=-\frac12$. However,
if we additionally drop the assumption of additive noise and instead
assume multiplicative noise, which depends on the distance to the
critical transition point, then the two generic 1-parameter scaling
exponents can decrease or increase as demonstrated in
\cite[Sec.~7.5]{KuehnCT2}. This is one scenario which could potentially 
account for the increased exponent  observed in Figure \ref{fig3}(B).\medskip 

Although the abstract mathematical analysis can be justified rigorously, 
it has to rely on a number of assumptions to achieve these results, 
including {e.g.}~smoothness of the ODEs \cite{KuehnScaleSN}, coupling of 
the normal form variable to the measured variable \cite[Sec.~7.4]{KuehnCT2}, 
weak-noise regime in comparison to parameter drift \cite[p.~96]{BerglundGentz}
and additive noise (as discussed above). The current state of abstract
social network modeling does not suffice to check all these assumptions
for a large realistic model. Interestingly, it has been shown in the context
of classical coarse-grained epidemic models that these assumptions may hold in
many cases; notably, epidemic models could potentially share many features
with social networks, see \cite[Sec.~7.2]{KuehnCT2} and \cite{OReganDrake}. 
In any case, variance and autocorrelation are relatively easy to measure using
a sliding window approach outlined in the previous section. In conclusion, this
provides the  mathematical motivation for our basic hypothesis that we would like
to test for in this study: Are there \textit{any} signs of increasing variance
and autocorrelation before spikes/jumps in network activity?

\section{Further Examples of Time Series}
\label{sec:further}
 
\begin{figure}
\includegraphics[width=\textwidth]{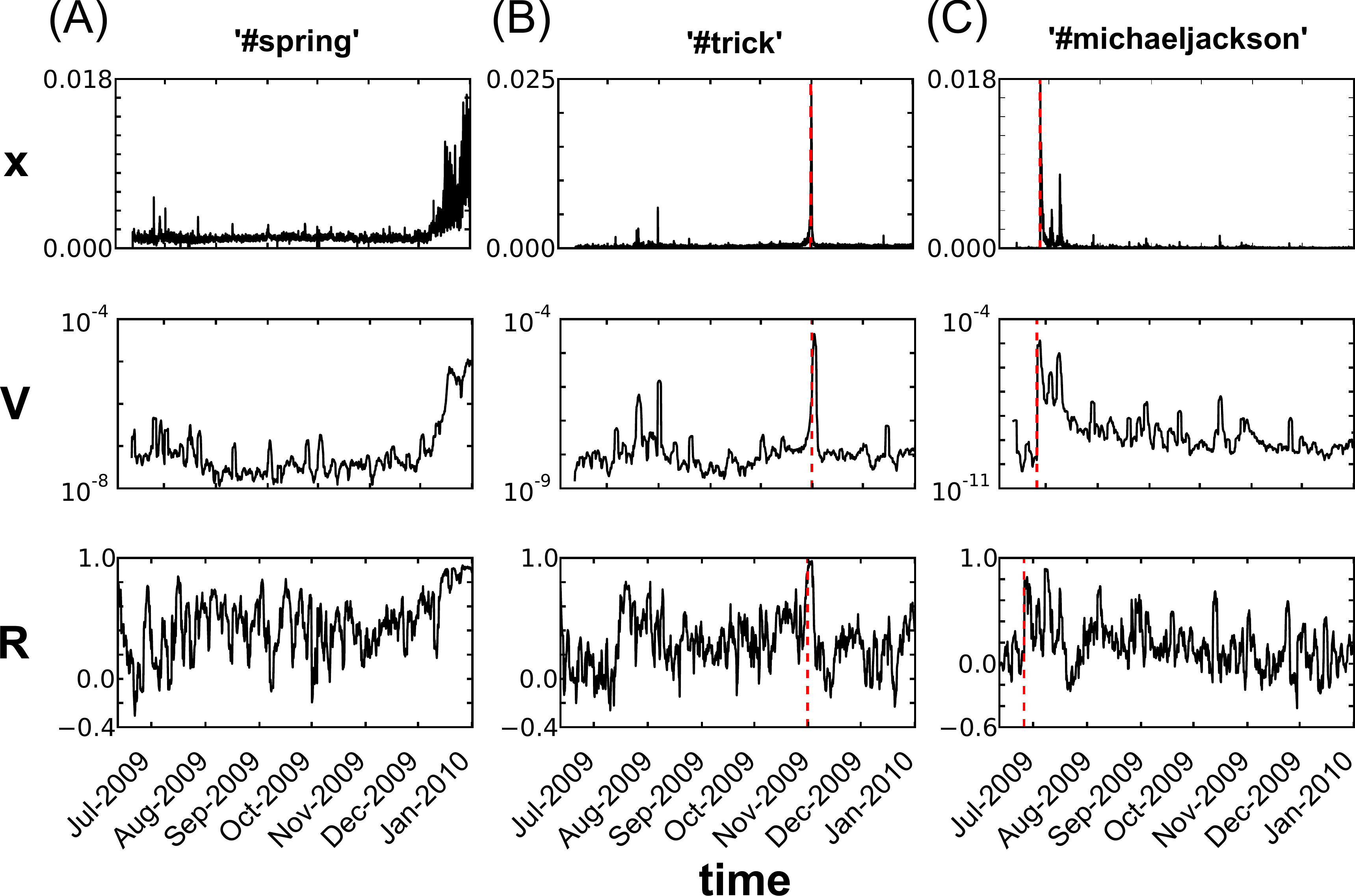}
\caption{\label{fig4}Further examples of time series with large-spike events.
Columns display from top to bottom normalized word frequency $x$, variance $V$
and lag-1 autocorrelation $R$ calculated from the time series of word frequency. 
(A) \#spring is another example of a time series with clear early warning signs;
in fact, the event ('spring') is again a-priori known, but by its own nature, it
is not an as  well-defined singular event as the examples shown in Figure~\ref{fig1}. 
(B) \#trick shows a larger spike which is difficult to predict and thus may evade
the time series analysis discussed here. (C) The sudden death (date marked) of
Michael Jackson  illustrates an example of a sudden time series spike that cannot
be anticipated due to its inherently exogenous nature.}
\end{figure}

\begin{figure}
\includegraphics[width=\textwidth]{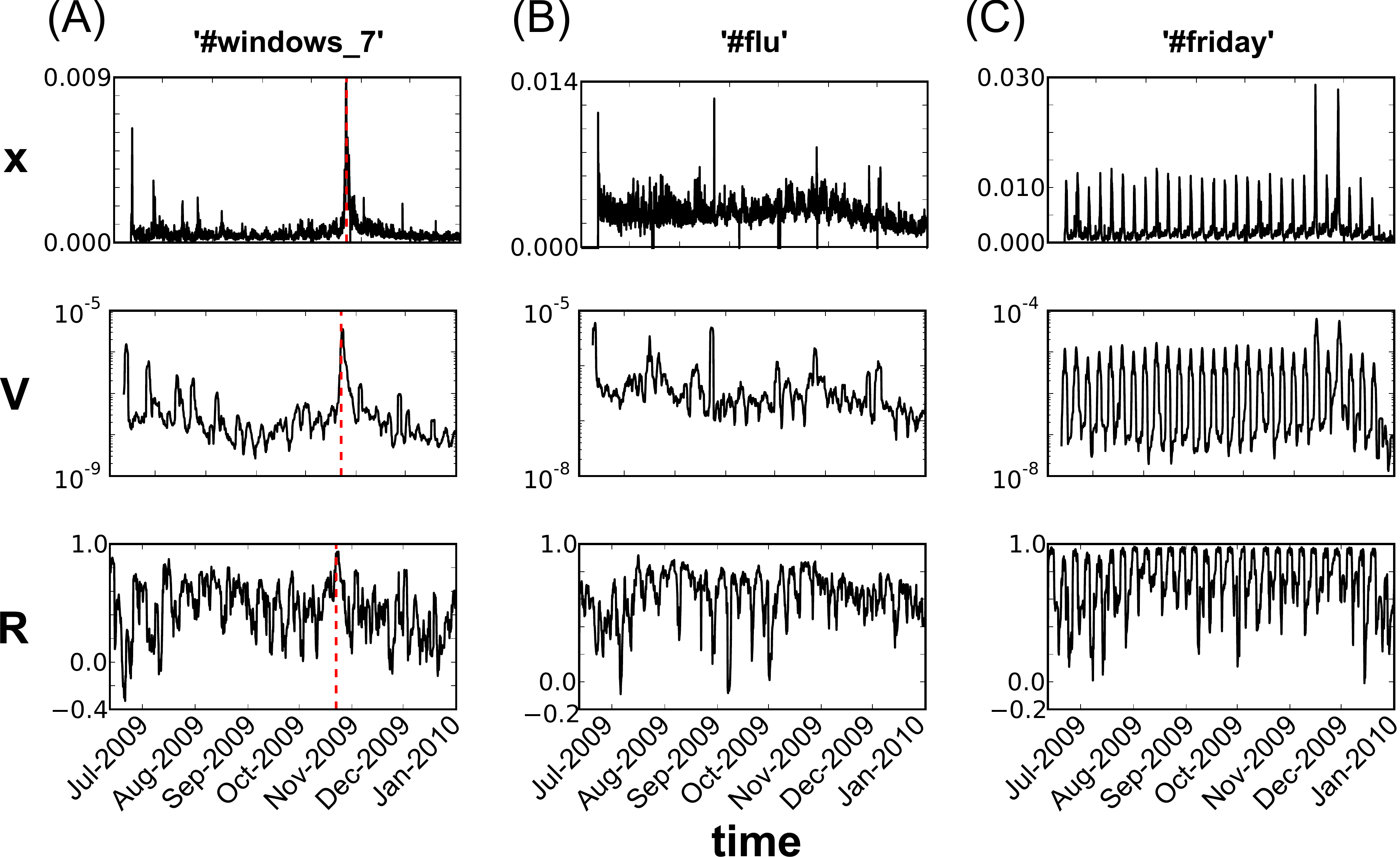}
\caption{\label{fig5}Further examples of time series with large-spike events.
Columns display from top to bottom normalized word frequency $x$, variance $V$
and lag-1 autocorrelation $R$ calculated from the time series of word frequency. 
(A) The red vertical dashed line indicates the a-priori known date of the
release event of the software package Windows 7. (B) The time series for
\#flu displays high noise levels where occasional larger spikes cannot be
predicted. (C) The time series for \#friday exhibits clear periodic driving
for self-evident reasons.}
\end{figure}

To illustrate various other phenomena that can be found in large-scale
social media, time series, we discuss a few additional hashtag time series,
shown in Figures~\ref{fig4} and \ref{fig5}. Another example exhibiting warning
signs for an a-priori known event is the time series for \verb|#spring| in
Figure \ref{fig4}(A). However, note that the event of 'spring' is not confined
to a single particular date and therefore, we observe a broad regime of large
activity instead of one distinct spike. Figure~\ref{fig4}(B) shows the time
series for \verb|#trick|, which has a very large peak at Halloween (marked
by a dashed red line), apparently due to its relation to the phrase
``trick-or-treat''; but the variance and autocorrelation begin to increase
only very close to the spike in the time series. This illustrates how indirect
indicators of an event could be substantially worse predictors than the hashtags
for the event themselves. Figure \ref{fig4}(C) shows the hashtag
\verb|#michaeljackson| exhibiting an extremely large spike at the day of
the sudden, unexpected death of Michael Jackson, but -- obviously -- without 
any warning signs preceding the event. The cause of this event is clearly
exogenous in nature  (in terms of social network dynamics) and thus, it is
impossible to detect (or anticipate, for that matter) early warning signs in
these time series data. 

Figure \ref{fig5}(A) shows the time series for the hashtag \verb|#windows_7|
related to the release event (red dashed line) of the software package Windows 7.
There is increasing variance as well as autocorrelation before the (a-priori known)
event. However, the warning signs are not as clear-cut as in the examples shown
in Figure~\ref{fig1}. In fact, the prediction time window seems rather short.
Thus, this example represents a case which lies in between the realm of predictable
and unpredictable events, at least as long as the detection of warning signs is
based upon the variance and autocorrelation analysis. Figure~\ref{fig5} shows
the time series for the hashtag \verb|#flu| which is very noisy. Although there
are several spikes in activity, these spikes appear clearly unpredictable as
-- from a modeling point of view -- seem to be noise-induced events. Finally,
Figure~\ref{fig5} shows a periodically-driven time series for \verb|#friday|.
Warning signs could potentially be possible to identify during each week before 
the spikes, but we do not examine this case in further detail since, to the best
of our knowledge, there is basically no theoretical background for warning signs
for systems with short rapid periodic forcing.


\begin{thebibliography}{10}

\bibitem{AgarwalLimWigand}
N.~Agarwal, M.~Lim, and R.~Wigand.
\newblock Raising and rising voices in social media - a novel methodological
  approach in studying cyber-collective movements.
\newblock {\em Business Inf. Syst. Eng.}, 4(3):113--126, 2012.

\bibitem{AshwinWieczorekVitoloCox}
P.~Ashwin, S.~Wieczorek, R.~Vitolo, and P.~Cox.
\newblock Tipping points in open systems: bifurcation, noise-induced and
  rate-dependent examples in the climate system.
\newblock {\em Phil. Trans. R. Soc. A}, 370:1166--1184, 2012.

\bibitem{BerglundGentz}
N.~Berglund and B.~Gentz.
\newblock {\em Noise-Induced Phenomena in Slow-Fast Dynamical Systems}.
\newblock Springer, 2006.

\bibitem{BimberFlanaginStohl}
B.~Bimber, A.J. Flanagin, and C.~Stohl.
\newblock Reconceptualizing collective action in the contemporary media
  environment.
\newblock {\em Comm. Theor.}, 15(4):365--388, 2005.

\bibitem{CarpenterBrock}
S.R. Carpenter and W.A. Brock.
\newblock Rising variance: a leading indicator of ecological transition.
\newblock {\em Ecology Letters}, 9:311--318, 2006.

\bibitem{Carpenteretal1}
S.R. Carpenter, J.J. Cole, M.L. Pace, R.~Batt, W.A. Brock, T.~Cline, J.~Coloso,
  J.R. Hodgson, J.F. Kitchell, D.A. Seekell, L.~Smith, and B.~Weidel.
\newblock Early warning signs of regime shifts: a whole-ecosystem experiment.
\newblock {\em Science}, 332:1079--1082, 2011.

\bibitem{Chertkovetal}
M.~Chertkov, S.~Backhaus, K.~Turzisyn, V.~Chernyak, and V.~Lebedev.
\newblock Voltage collapse and {ODE} approach to power flows: analysis of a
  feeder line with static disorder in consumption/production.
\newblock {\em arXiv:1106.5003v1}, pages 1--8, 2011.

\bibitem{CimatoribusDrijfhoutLivinavanderSchrier}
A.A. Cimatoribus, S.S. Drijfhout, V.~Livina, and G.~van~der Schrier.
\newblock {Dansgaard-Oeschger} events: tipping points in the climate system.
\newblock {\em Climate of the Past Discussion}, 8(5):4269--4294, 2012.

\bibitem{Cohen}
R.~Cohen.
\newblock Facebook and {Arab} dignity.
\newblock {\em New York Times}, 2011.
\newblock {Retrieved from
  http://www.nytimes.com/2011/01/25/opinion/25iht-edcohen25.html on 4 November 2013}.

\bibitem{DaiVorselenKorolevGore}
L.~Dai, D.~Vorselen, K.S. Korolev, and J.~Gore.
\newblock Generic indicators for loss of resilience before a tipping point
  leading to population collapse.
\newblock {\em Science}, 336:1175--1177, 2012.

\bibitem{Dakosetal1}
V.~Dakos, E.H. van Nes, R.~Donangelo, H.~Fort, and M.~Scheffer.
\newblock Spatial correlation as leading indicator of catastropic shifts.
\newblock {\em Theor. Ecol.}, 3(3):163--174, 2009.

\bibitem{DitlevsenJohnsen}
P.D. Ditlevsen and S.J. Johnsen.
\newblock Tipping points: early warning and wishful thinking.
\newblock {\em Geophys. Res. Lett.}, 37:19703, 2010.

\bibitem{DrakeGriffen}
J.M. Drake and B.D. Griffen.
\newblock Early warning signals of extinction in deteriorating environments.
\newblock {\em Nature}, 467:456--459, 2010.

\bibitem{Gladwell}
M.~Gladwell.
\newblock {\em The Tipping Point: How Little Things Can Make a Big Difference}.
\newblock Back Bay Books, 2002.

\bibitem{GH}
J.~Guckenheimer and P.~Holmes.
\newblock {\em Nonlinear Oscillations, Dynamical Systems, and Bifurcations of
  Vector Fields}.
\newblock Springer, New York, NY, 1983.

\bibitem{HudsonFlannes}
L.~Hudson and M.~Flannes.
\newblock {The Arab Spring: Anatomy} of a tipping point.
\newblock {\em Aljazeera}, 2011.
\newblock {Retrieved from
  http://www.aljazeera.com/indepth/opinion/2011/08/201183081433165611.html
   on 4 November 2013}.

\bibitem{KuehnScaleSN}
C.~Kuehn.
\newblock {Scaling of saddle-node bifurcations: degeneracies and rapid
  quantitative changes}.
\newblock {\em J. Phys. A: Math. and Theor.}, 42(4):045101, 2009.

\bibitem{KuehnCT1}
C.~Kuehn.
\newblock {A mathematical framework for critical transitions: bifurcations,
  fast-slow systems and stochastic dynamics}.
\newblock {\em Physica D}, 240(12):1020--1035, 2011.

\bibitem{KuehnCT2}
C.~Kuehn.
\newblock {A mathematical framework for critical transitions: normal forms,
  variance and applications}.
\newblock {\em J. Nonlinear Sci.}, 23(3):457--510, 2013.

\bibitem{Kuznetsov}
Yu.A. Kuznetsov.
\newblock {\em Elements of Applied Bifurcation Theory}.
\newblock Springer, New York, NY, 3rd edition, 2004.

\bibitem{Lamberson}
P.J. Lamberson and S.E. Page.
\newblock {Tipping Points}
\newblock {\em International Quarterly Journal of Political Science}, 7(2):175--208, 2012.

\bibitem{LadeGross}
S.J. Lade and T.~Gross.
\newblock Early warning signals for critical transitions: a generalized
  modeling approach.
\newblock {\em PLoS Comp. Biol.}, 8:e1002360--6, 2012.

\bibitem{Lenton}
T.M. Lenton.
\newblock Early warning of climate tipping points.
\newblock {\em Nature Climate Change}, 1(4):201--209, 2011.

\bibitem{Lim1}
M.~Lim.
\newblock {Tahrir Square} was a foreseeable surprise: tracing the history of
  {Egyptian} online activism.
\newblock {\em Slate}, 2011.
\newblock {Retrieved from
  http://www.slate.com/}{articles/}\\{technology/}{future\_tense}{/2011/07/}{tahrir\_square\_}{was\_a\_foreseeable}{\_surprise.html} on 4 November 2013.

\bibitem{Lim}
M.~Lim.
\newblock Clicks, cabs, and coffee houses: social media and oppositional
  movements in {Egypt}, 2004-2011.
\newblock {\em J. Communication}, 62(2):213--248, 2012.

\bibitem{Lotanetal}
G.~Lotan, E.~Graeff, M.~Ananny, D.~Gaffney, I.~Pearce, and D.~Boyd.
\newblock The revolutions were tweeted: {Information} flows during the 2011
  {Tunisian and Egyptian} revolutions.
\newblock {\em Int. J. Commun.}, 5:1375--1405, 2011.

\bibitem{MeiselKuehn}
C.~Meisel and C.~Kuehn.
\newblock On spatial and temporal multilevel dynamics and scaling effects in
  epileptic seizures.
\newblock {\em PLoS ONE}, 7(2):e30371, 2012.

\bibitem{OReganDrake}
S.M. O'Regan and J.M. Drake.
\newblock Theory of early warning signals of disease emergence and leading
  indicators of elimination.
\newblock {\em Theor. Ecol.}, 6(3):333--357, 2013.

\bibitem{RomeroMeederKleinberg}
D.~Romero, B.~Meeder, and J.~Kleinberg.
\newblock Differences in the mechanics of information diffusion across topics:
  {Idioms,} political hashtags, and complex contagion on {Twitter}.
\newblock In {\em Proceedings of the 20th International Conference on World
  Wide Web}, pages 695--704, 2011.

\bibitem{Schefferetal}
M.~Scheffer, J.~Bascompte, W.A. Brock, V.~Brovkhin, S.R. Carpenter, V.~Dakos,
  H.~Held, E.H. van Nes, M.~Rietkerk, and G.~Sugihara.
\newblock Early-warning signals for critical transitions.
\newblock {\em Nature}, 461:53--59, 2009.

\bibitem{Slater}
D.~Slater.
\newblock Early warning signals of tipping-points in blog posts.
\newblock {\em preprint}, 2012.
\newblock {http://www.mitre.org/work/tech\_papers/2012/12\_4711/12\_4711.pdf} on 4 November 2013.

\bibitem{ThompsonSieber2}
J.M.T. Thompson and J.~Sieber.
\newblock Climate tipping as a noisy bifurcation: a predictive technique.
\newblock {\em IMA J. Appl. Math.}, 76(1):27--46, 2011.

\bibitem{TufekciWilson}
Z.~Tufekci and C.~Wilson.
\newblock Social media and the decision to participate in political protest:
  observations from {Tahrir Square}.
\newblock {\em J. Communication}, 62(2):363--379, 2012.

\bibitem{Venegasetal}
J.G. Venegas, T.~Winkler, G.~Musch, M.F.~Vidal Melo, D.~Layfield,
  N.~Tgavalekos, A.J. Fischman, R.J. Callahan, G.~Bellani, and R.S. Harris.
\newblock Self-organized patchiness in asthma as a prelude to catastrophic
  shifts.
\newblock {\em Nature}, 434:777--782, 2005.

\bibitem{Veraartetal}
A.J. Veraart, E.J. Faassen, V.~Dakos, E.H. van Nes, M.~Lurling, and
  M.~Scheffer.
\newblock Recovery rates reflect distance to a tipping point in a living
  system.
\newblock {\em Nature}, 481:357--359, 2012.

\bibitem{Wangetal}
R.~Wang, J.A. Dearing, P.G. Langdon, E.~Zhang, X.~Yang, V.~Dakos, and
  M.~Scheffer.
\newblock Flickering gives early warning signals of a critical transition to a
  eutrophic lake state.
\newblock {\em Nature}, 492:419--422, 2012.

\bibitem{Webster}
S.~Webster.
\newblock Has social media revolutionized revolutions?
\newblock {\em World News}, 87(15), 2011.
\newblock {Retrieved from
  http://www.jcunews.com/2011/02/16/has-social-media-revolutionized-revolutions/}.

\bibitem{YangLeskovec}
J.~Yang and J.~Leskovec.
\newblock Patterns of temporal variation in online media.
\newblock In {\em Proceedings of the fourth ACM International Conference on Web
  Search and Data Mining}, pages 177--186, 2011.

\end{thebibliography}
\end{document}